\documentclass[prd, showpacs]{revtex4}
\usepackage{graphicx}
\usepackage{amssymb}
\usepackage{dcolumn}
\usepackage{bm}
\newcommand{\cc}{\cite}

\newcommand{\be}{\begin{equation}}
\newcommand{\ee}{\end{equation}}

\def\ve{\varepsilon}
\def\w{\omega}

\def\pd{\partial}
\def\f{\phi}
\def\F{\Phi}

\def\L{\Lambda}

\def\ex{\hbox{e}}

\def\F{\Phi}
\def\<{\langle}
\def\>{\rangle}

\def\a{\alpha}
\def\b{\beta}
\def\g{\gamma}  \def\G{\Gamma}
\def\d{\delta}  \def\D{\Delta}
\def\l{\lambda}   \def\L{\Lambda}
\def\s{\sigma}
\def\r{\rho}  
\def\x{\xi}
\def\c{\chi}

\def\m{\mu}
\def\n{\nu}

\def\w{\omega}

\def\v{\vec}

\def\vf{\varphi}
\def\({\left(}
\def\[{\left[}
\def\){\right)}
\def\]{\right]}
\def\coth{\hbox{coth}}

\def\pd{\partial}
\def\dk{{d^n k \over (2\pi)^n}}
\def\Tr{\hbox{Tr}}
\def\pa{{\cal P}}
\def\w1{W^{(1)}}
\def\v1{V^{(1)}}
\def\prop{D_{\m\n}}

\def\dI{\int \! dn(\r)}

\begin{document}
\title{Instanton corrections to quark form factor \protect \\ at large momentum transfer}
\author{Alexander E. Dorokhov}
\email{dorokhov@thsun1.jinr.ru}
\author{Igor O. Cherednikov}
\altaffiliation[Also at:]{
Institute for Theoretical Problems of Microphysics,
Moscow State University, 119899 Moscow, Russia} \email{igorch@thsun1.jinr.ru}
\affiliation{
Bogoliubov Laboratory of Theoretical Physics,
Joint Institute for Nuclear Research,
141980 Dubna, Russia}
\date{\today}
\begin{abstract}
Within the Wilson integral formalism,
we discuss the structure of nonperturbative corrections to the quark form factor
at large momentum transfer analyzing the infrared renormalon and instanton effects.
We show that the nonperturbative effects determine the initial value for the
perturbative evolution of the quark form factor and attribute their general structure
to the renormalon ambiguities of the perturbative series.
It is demonstrated that the instanton contributions result in the finite renormalization
of the next-to-leading perturbative result and numerically are characterized by a small
factor reflecting the diluteness of the QCD vacuum within the instanton liquid model.
\end{abstract}
\maketitle
\section{\label{sec:level1}Introduction}

The various aspects of the instanton induced effects in high energy QCD processes
had been addressed at the very beginning of the instanton era (see, {\it e.g.}, Ref. 
\cc{EL}), and this study has been continued in later decade
\cc{BAL}. Recently, the interest in them has been revived \cc{Sh, KKL, RW, SRP, DCH},
and the hope of the direct detection of the instanton induced effects has
appeared \cc{RH}. One of the important questions in the description of hadronic exclusive
processes is the behavior of the form factors in various energy
domains. The present paper is devoted to the analysis of the corrections to the 
elastic quark form factor
at large momentum transfer induced by the infrared (IR) renormalon \cc{REN} and instanton
effects, treating the latter within the framework of the instanton liquid model of 
the QCD vacuum \cc{REV}.

From the theoretical point of view, the form factor analysis requires a perturbative
resummation procedure beyond the standard renormalization group techniques, since it
exhibits a double-logarithmic behavior. In addition to this, 
the resummation techniques developed for this
particular case can be applied to the study of many other processes which possesses
the logarithmic enhancements near the kinematic boundaries.
On the other hand, in addition to this obvious theoretical use, the computation of the quark
form factor has important phenomenological applications. A similar resummation approach is also
used in study of the near-forward quark-quark scattering and evaluation of the
soft Pomeron properties where the nonleading logarithmic contributions are quite important
\cc{KRP}. The application of this formalism in heavy quark effective
theory can be also useful \cc{KRH}. The quark form factor itself enters into
the various cross sections of high-energy processes \cc{HARD}. In particular, this quantity
finds the most straightforward phenomenological application in the
total cross section of the Drell-Yan process in the deep inelastic scattering (DIS) scheme, 
which is proportional to the ratio
of the timelike and spacelike form factors \cc{PAR, STER, MAG}. In this case, the exponentiated
quark form factor is expressed in terms of the
evolution equation, and can be evaluated, in principle, to any order in perturbation theory.
The analysis shows that this is the high-energy asymptotic behavior that is important, and
all the logarithmic contributions must be taken into account equally, while the power corrections
could be neglected \cc{MAG}. The investigation of the electomagnetic quark form factors
in moderate and low-energy domains can shed light on the problem of scaling violation in DIS and
the structure of constituent quarks \cc{SCAL}.

The color singlet quark form factor is determined via the elastic scattering
amplitude of a quark in electromagnetic field
\be
{\cal M}_\m =F_q\[(p_1-p_2)^2\]  \bar u(p_1) \g_\m v(p_2) \ \ ,
\ee where $u(p_1), \ v(p_2)$ are the spinors of outgoing and incoming
quarks.
The kinematics of the process is described in terms of the scattering
angle $\c$:
\be \cosh \c = {(p_1 p_2) \over m^2} = 1 + {Q^2 \over 2 m^2} \ \ ,
\ \ Q^2 = - (p_2 - p_1)^2>0 \ \ , \ \ p_1^2=p_2^2 =m^2  \ .\label{m1} \ee

It is known that the leading large-$Q^2$ asymptotics of the quark
form factor is given by the exponentiation of the one-loop term \cc{DL}
\be
F^{(1)}_q(Q^2) = \exp\(-{\a_s \over 4\pi} C_F \ln^2{Q^2 \over \l^2} \)
+ O\(\a_s^n \ln^{2n-1}{Q^2 \over \l^2}\) \ , \label{dl}
\ee where $\l$ is an IR cutoff parameter.
In general, for a  correct consideration of the non-leading asymptotic contributions one has to
resum all perturbative (such as $O\(\a_s^n \ln^{2n-1}{Q^2}\)$, $O\(\a_s^n \ln^{2n-2}{Q^2}\)$,
{\it etc})
as well as nonperturbative terms. An effective framework for resummation of perturbative and nonperturbative
contributions is
provided by the Wilson integral approach \cc{NACH}.
Within this framework, the resummation of all logarithmic terms
coming from the soft gluon subprocesses
allows us to express the quark form factor (\ref{dl}),
in terms of the vacuum average of the gauge invariant path ordered
Wilson integral \cc{MMP}
\be
W (C_\c)  = {1 \over N_c} \Tr \<0|  \pa \exp \( i g \int_{C_\c} \! d x_{\m} \hat A_{\m}
(x) \)|0\> \ . \label{1a} \ee
In Eq. (\ref{1a}) the integration path corresponding to considering process
goes along the closed contour $C_\c$: the angle (cusp) with infinite sides.
The gauge field
\be
\hat A_{\m} (x) = T^a A^a_{\m}(x)\ , \   \ T^a = {\lambda^a
\over 2} \ \ ,  \ee belongs to the Lie algebra of the gauge group
$SU(N_c)$, while the Wilson loop operator $\pa \ex^{ig\int\! dx A(x)}$ lies in
its fundamental representation.

In our recent paper \cc{DCH}, we applied the Wilson integral formalism to
evaluation of the perturbative and nonperturbative contributions to the
color singlet quark form factor at the low normalization point $\m$ of order
of the inverse instanton size within the instanton liquid model. In the
present work, considering the renormalization group (RG) evolution equation we extend the analysis
to the limit of large momentum transfers focusing on the
asymptotic behavior. We show that the nonperturbative effects determine the initial
value for the perturbative evolution, find their general structure by
analyzing the renormalon ambiguities of the perturbative series, and establish the correspondence
between them and the instanton induced contribution.

The paper is organized as follows. 
In Sec. II we reproduce the known results of the perturbative one-loop
calculation, and derive the evolution equations taking into account the
nonperturbative contribution as the initial value for perturbative
evolution. In Sec. III,
we study the consequences of the IR renormalon ambiguities of the perturbative series and
show how the latter prescribes the form of the nonperturbative corrections to the asymptotic behavior
of the form factor at large $Q^2$.
In Sec. IV, these nonperturbative effects are estimated
in the weak-field approximation within the instanton model of the QCD vacuum.
Finally, the large $Q^2$ behavior of the form factor is analyzed taking into account the leading
perturbative, IR renormalon, and instanton induced contributions. The latter
are found to be determined by small factor expressed via the
parameters of the instanton liquid model.

\section{Analysis of the perturbative contributions to the Wilson integral}

The Wilson integral (\ref{1a}) can be presented as a series
$$
W(C_\c)  = 1 + {1 \over N_c}\<0| \sum_{n=2} \ (ig)^n \int_{C_\c}\int_{C_\c} ...\int_{C_\c}
 \! dx_{\m_n}^n \ dx_{\m_{n-1}}^{n-1}... dx_{\m_1}^1 \cdot $$
\be \cdot \ \theta (x^n, x^{n-1}, ... , x^1)
\ \Tr \[\hat A_{\m_n} (x^n) \hat A_{\m_{n-1}} (x^{n-1})... \hat A_{\m_{1}} (x^1) \]|0\>
\ , \label{expan1} \ee where the function $\theta (x)$ orders the color matrices along
the integration contour.
In the present work, we restrict ourselves with the study of the  leading order
(one loop---for the perturbative gauge field and weak-field limit for the
instanton) terms of the expansion (\ref{expan1}) which are given by the
expression
\be
\w1 (C_\c) =  -{g^2 C_F \over 2}
\ \int_{C_\c}\! dx_\m \int_{C_\c}\! dy_\n \ \prop (x-y)
\ , \label{g1} \ee where the gauge field propagator $\prop(z)$ in
$n$-dimensional space-time $(n = 4 - 2 \ve)$ can be presented in the form
\be
\prop(z) = g_{\m\n}
\pd_z^2 \D_1(\ve, z^2, \m^2/\l^2) - \pd_\m\pd_\n \D_2(\ve, z^2, \m^2/\l^2) \ .
\label{st1} \ee
The exponentiation theorem for non-Abelian path-ordered Wilson integrals
\cc{KR, ET} allows us to express (to one-loop accuracy) the Wilson integral (\ref{1a})
as the exponentiated one-loop term of the series (\ref{expan1}):
\be
W(C_\c) = \exp\[\w1(C_\c) + O(\a_s^2)\] \ . \label{eq:expn1}
\ee

In general, the expression (\ref{g1}) contains ultraviolet (UV) and IR divergences, that
can be multiplicatively renormalized in a consistent way \cc{BRA}.
In contrast to the previous paper \cc{DCH}, we use the dimensional
regularization in order to work with UV singularities, and define the ``gluon mass''
$\l^2$ as the IR regulator and the parameter $\mu^2$ as the UV normalization point.
The dimensionally regularized formula for the leading order (LO) term (\ref{g1})
can be written as \cc{DCH}
\be
\w1 (C_\c; \ve, \m^2/\l^2, \a_s)
= 8 \pi \a_s  C_F h (\c) (1 - \ve)  \D_1(\ve, 0, \m^2/\l^2) \ ,
 \label{pe1} \ee
where $h(\c)$ is the universal cusp factor, 
\be
h(\c) = \c \coth \c -1 \ ,  \ee
and, in case of the perturbative field, 
\be
\D_1 (\ve, 0, \m^2/\l^2) = - {1 \over 16\pi^2} \(4\pi {\m^2 \over
\l^2}\)^{\ve} \ {\G(\ve) \over 1 - \ve} \ .  \label{eq:pr01}
\ee
The independence of the expression (\ref{pe1}) of the function $\D_2$ is a
direct consequence of the gauge invariance.
Then, in the one-loop approximation,
\be
W(C_\c; \ve, \m^2/\l^2, \a_s(\mu)) = 1 - {\a_s(\m) \over 2\pi} C_F h(\c) \({1 \over \ve}
- \g_E + \ln 4\pi + \ln{\m^2\over\l^2}
\),
\ee and the cusp dependent renormalization constant \cc{BRA}, within the
modified minimal subtraction scheme, reads
\be
Z_{cusp} \[C_\c; \ve, \a_s(\m)\] =
1 + {\a_s(\m) \over 2\pi} C_F h(\c) \({1 \over \ve} - \g_E + \ln 4\pi \) \
.
\ee
The detailed description of the renormalization procedure
within the present approach has been made in \cc{DCH} and will be omitted
here for brevity.

Using Eq. (\ref{pe1}), one finds the known one-loop result
for the perturbative field, which  contains the dependence
on the UV normalization point $\m^2$ and IR cutoff $\l^2$
({\it e.g.}, \cc{KR}):
\be \w1_{pt} (C_\c)
=  - {\a_s (\m) \over 2 \pi} C_F  h(\c) \ln {\m^2 \over \l^2}
\ . \label{5} \ee
Therefore, in the leading order the kinematic dependence of the expression (\ref{g1}) is
factorized into the function $h(\c)$, which at large-$Q^2$ is approximated by
\be h(\c) \propto \ln {Q^2 \over m^2} \ . \label{eq:lar1} \ee
In this regime, the dependence of $W$ on the UV normalization scale $\m$ (which can also be treated
as an arbitrary factorization scale dividing the hard and soft subprocesses 
\cc{KRC1}) is governed by the renormalization group (RG) equation
\be
\(\m{\pd \over \pd \m} + \b(g) {\pd \over \pd g} \)
{d \  \ln \ W (Q^2) \over d \ \ln Q^2}
 = - \G_{cusp} \[ \a_s(\m) \] \ , \label{2}
\ee where $\G_{cusp}(\a_s)$ is the universal cusp anomalous dimension evaluated
in the perturbation theory. In Eq. (\ref{2}), we take the logarithmic
derivative in $Q^2$ in order to avoid problems with light-cone singularities
at $m^2=0$ \cc{KRC1}.
The solution of the RG equation leads to the evolution equation
\be
{d \ \ln W (Q^2) \over d \ \ln Q^2} = -
\int_{\l^2}^{\m^2} \! {d\x \over 2\x} \ \G_{cusp}\[ \a_s(\x) \] +
{d \  W_{np} (Q^2) \over d \ \ln Q^2} \ , \label{inc1}
\ee where the function $W_{np}$ gives the initial condition at $\m^2 = \l^2$
and has to be found by the nonperturbative methods \cc{KRREN, TAF}. Solving Eq.
(\ref{inc1}), we take the arbitrary upper bound for the squared momenta of soft gluons
equal to the hard scale $\m^2 = Q^2$ and find
\be
\ln \ {W (Q^2) \over W(Q_0^2)} = - \int^{Q^2}_{Q_0^2}\! {d x \over x}\[
\ \int_{\l^2}^{x} \! {d \x \over 2
\x}\ \G_{cusp}\[ \a_s(\x)\] - {d \  W_{np} (x) \over d \ \ln  x} \] \ ,
\label{eq:ev1}
\ee which immediately leads to the
conclusion that the leading large-$Q^2$ behavior of the quark form factor $F_q(Q^2)$
including all logarithmic corrections is controlled by
the universal cusp anomalous dimension (\ref{2}) and can be expressed in
the following form (for comparison, see \cc{KRC1}):
$$
F_q(Q^2) = W(Q^2) $$ \be = \exp\[-\int_{Q_0^2}^{Q^2}\! {d\x \over 2\x} \  \ln{Q^2 \over \x}
\ \G_{cusp} \[ \a_s(\x) \]  - \ln {Q^2 \over Q_0^2}\ \int_{\l^2}^{Q_0^2} \! {d\x \over
2\x} \ \G_{cusp}\[ \a_s(\x) \] +  W_{np} (Q^2) \] W_0 \ , \label{1} \ee
where $W_0 = W(Q_0^2)$ contains both perturbative and nonperturbative
contributions. From the one-loop result (\ref{5}),
the cusp anomalous dimension which satisfies the RG equation
(\ref{2}) in one-loop order is given by:
\be {\G_{cusp}^{(1)}} ( \a_s (\m)) = {\a_s (\m) \over \pi} C_F \ .  \label{cp} \ee
Substituting the anomalous dimension (\ref{cp}) in the one-loop approximation for the strong
coupling into the Eq. (\ref{1}), one
finds
$$
F_q^{(1)}(Q^2)  $$ \be = \exp \[ - {2C_F \over \b_0} \(\ln {Q^2 \over \L^2}
\ln{\ln(Q^2/\L^2) \over \ln(Q_0^2/\L^2)} - \ln {Q^2 \over Q_0^2}
\(1 - \ln{\ln(Q_0^2/\L^2) \over \ln(\l^2/\L^2)} \) \) +  W_{np} (Q^2) \]
F^{(1)}(Q_0^2)\ , \label{npc0} \ee where $\L$ is the QCD scale. The singularity in Eq. (\ref{npc0})
originates from the region where the IR cutoff approaches $\L$, {\it
i. e., } where the coupling constant $\a_s$ increases, and then may have a
nonperturbative nature.

\section{Effects of the IR renormalons}

In order to determine the structure of the nonperturbative function $W_{np}$ in
Eqs. (\ref{1}, \ref{npc0}), it is instructive to study the
corrections due to IR renormalons \cc{REN}. In the present situation, one can expect
the corrections proportional to the powers of both scales $\m^2$ and
$\l^2$. However, taking into account the evolution in $\m^2$ to the hard
characteristic scale of the process $Q^2$ (\ref{eq:ev1}), we treat the power $\m^2$ terms
to be strongly suppressed, and focus on the power $\l^2$ corrections.
To find them, let us consider the perturbative function 
$\D_1(\ve, 0, \m^2/\l^2)$ in the Eq. (\ref{pe1}).
The insertion of the fermion bubble 1-chain to the one-loop order expression
(\ref{g1}) is equivalent to replacement of the frozen coupling constant $g^2$ by the
running one $g^2 \to g^2 (k^2) = 4\pi \a_s(k^2)$
\cc{KRREN}:
\be \widetilde\D_1(\ve, 0, \m^2/\l^2) = - 4\pi \m^{2\ve}  \int \! \dk \a_s(k^2){\ex^{ikz} \delta(z^2)
 \over k^2(k^2+\l^2)}\ .
\label{ren1} \ee
For the sake of convenience, we work here in Euclidean space.
Using the integral representation for the one-loop running coupling $\a_s(k^2) = \int_0^\infty \!
d\s (\L^2/k^2)^{\s b }$, $b = \b_0 /4\pi$, we find
\be
\widetilde\D_1(\ve, 0, \m^2/\l^2) = - {1 \over \b_0(1-\ve)} \(4\pi {\m^2 \over \l^2}\)^\ve \int_0^{\infty} \!
dx\
{\G(1-x -\ve) \G(1+x+\ve) \over (x +\ve ) \G(1 - \ve )} \({\L^2 \over \l^2}\)^x \ . \label{gamma} \ee
To define properly the integral on the right hand sided of Eq. (\ref{gamma}), 
one needs to specify a prescription to go around the poles, which
are at the points $\bar x_n = n , \ n \in {\mathbb N}$. Of course, the result
of integration will depend on this prescription giving an ambiguity proportional
to $\(\L^2 / \l^2 \)^n$ for each pole. Then, the IR renormalons produce the power
corrections to the one-loop perturbative result, which we assume to exponentiate with the
latter \cc{KRREN, TAF}. Extracting from Eq. (\ref{gamma}) the UV singular part in vicinity of the
origin $x =0$, we divide the integration interval $[0, \infty]$ in two parts
$[0, \d]$ and $[\d, \infty]$, where $\d < 1$. This procedure allows us to
evaluate the ultraviolet and the renormalon-induced pieces separately. For the
ultraviolet piece, we apply the expansion of the integrand in
$\D_1$ in powers of small $x$ and replace the ratio of $\G$ functions by
$\exp(-\g_E \ve)$:
\be \widetilde\D_1^{UV} (\ve, 0, \m^2/\l^2) = -  {1 \over \b_0 (1- \ve)}
\sum_{k, n=0} (-)^n { \(  \ln 4\pi - \g_E + \ln {\m^2  \over \l^2}\)^k
\over k! \ve^{n-k+1}}  \
\int_0^{\d} \! dx \ x^{n} \ \(\L^2 \over \l^2 \)^x \ , \label{rnm1} \ee
which after subtraction of the poles in the modified minimal subtraction scheme becomes
\be
\widetilde\D_1^{UV} (0, \m^2/\l^2) =  {1 \over \b_0 ( 1- \ve)}\  \sum_{n=1}\(\ln {\m^2 \over \l^2}
\)^n {(-)^n \over n!} \ \int_0^{\d} \! dx x^{n-1} \(\L^2 \over \l^2 \)^x \ .
\label{eq:iks}\ee
In analogy with results of Ref. \cite{Mikh98}, this expression may be rewritten in a closed form as
\be
\widetilde\D_1^{UV} ( 0, \m^2/\l^2) =
{1 \over \b_0 (1- \ve)}
\int_{0}^{\delta}\frac{dx}{x}
\[\ex^{-x\ln\frac{\mu^{2}}{\L^{2}}}
-  \ex^{-x\ln\frac{\l^{2}}{\L^{2}}
}\].
\label{Dex}\ee
Substituting
\be
{d \ \w1 (Q^2) \over d \ \ln Q^2}= 2 C_F (1-\ve) \widetilde\D_1^{UV}(0, \m^2/\l^2)
\label{Dex1}\ee
into Eq. (\ref{2}) one finds
\be
\(\m{\pd \over \pd \m} + \b(g) {\pd \over \pd g} \)
{d \  \ln W^{(1)} (Q^2) \over d \ \ln Q^2}
 = - \G_{cusp}^{(1)} (\a_s(\m)) \(1-\exp{\[-\d\frac{4\pi}{\b_0\a_s(\m)}\]}\).
\ee The second exponent in the last equation yields the power suppressed
terms $\(\L^2/Q^2\)^\d$ in large-$Q^2$ regime.
In the leading logarithmic approximation (LLA) Eq. (\ref{Dex1}) is reduced to
\be
{d \ \w1 (Q^2) \over d \ \ln Q^2} = - {2C_F \over \b_0} \(
\ln{\ln(\m^2/\L^2) \over \ln(\l^2/\L^2)}  \) \ . \label{ptr2}
\ee
The last expression obviously satisfies the perturbative evolution
equation (\ref{2}).

The remaining integral in Eq. (\ref{gamma}) over the interval $[\d, \infty]$
is evaluated at $\ve =0$ since there are no UV singularities.
The resulting expression does not depend on the normalization point $\m$, and thus
it is determined by
the IR region including nonperturbative effects. It contains the renormalon ambiguities
due to different prescriptions in going around the poles $\bar x_n$ in the Borel
plane which yields the power corrections to the quark form factor.

After the substitution $\m^2 = Q^2$ and integration over $ d (\ln \ Q^2)$, we find
in LLA (for comparison, see Eq. (\ref{npc0}))
$$
F_q^{ren}(Q^2) $$ \be = \exp\[- {2C_F \over \b_0} \[\ln {Q^2 \over \L^2}
\ln{\ln(Q^2/\L^2) \over \ln(Q_0^2/\L^2)} - \ln {Q^2 \over Q_0^2}
\(1 - \ln{\ln(Q_0^2/\L^2) \over \ln(\l^2/\L^2)} \) \] -  \ln {Q^2 \over Q_0^2}
\f_{ren} (\l^2, \L^2) \] F^{ren}(Q_0^2)\ , \label{npc}
\ee
where the function $ \f_{ren}(\l^2, \L^2) = \sum_{k=0} \f_k (\L^2/\l^2)^k$
accumulates the effects of
the IR renormalons, as well as the other nonperturbative information. The coefficients
$\f_k $ cannot be calculated in perturbation theory and can be treated
as the minimal set of nonperturbative parameters.  It is worth noting that
the logarithmic $Q^2$ dependence of the renormalon induced corrections in the large-$Q^2$
regime is factorized, and thus the Eq. (\ref{npc}) reproduces
exactly the structure of nonperturbative contributions found in the
one-loop evolution equation (\ref{npc0}) with respect to the large-$Q^2$ asymptotic behavior.

\section{Large-$Q^2$ behaviour of the instanton induced contribution}

Let us consider the instanton induced corrections to the perturbative result. The instanton field
is given by
\be \hat A_\m (x; \r) = A^a_{\m} (x; \r) {\sigma^a \over 2} = {1 \over g}
 {\hbox{\bf R}}^{ab} \sigma^a {\eta^\pm}^b_{\m\n} (x-z_0)_\n \vf
(x-z_0; \r) , \label{if1}\ee
where ${\hbox{\bf R}}^{ab}$ is the color orientation matrix $\[ a = {1,..., (N_c^2-1)}, b=1,2,3 \]$,
$\s^a$'s are the Pauli matrices,
and $(\pm)$ corresponds to the instanton, or anti-instanton.
The averaging of the Wilson operator over the nonperturbative vacuum is reduced to the integration
over the coordinate of the instanton center $z_0$, the color orientation and the
instanton size $\r$.  The measure for the averaging over the instanton ensemble
reads $dI = d{\hbox{\bf R}} \ d^4 z_0 \ dn(\r) $, where
$ d{\hbox{\bf R}}$ refers to the averaging over color orientation 
and $dn(\r)$ depends on the choice of the instanton size distribution.
Taking into account Eq. (\ref{if1}),
we write the Wilson integral (\ref{1a}), which defines the instanton induced
contribution to the nonperturbative part in Eq. (\ref{1}), in the single instanton approximation
in the form
\be
W_I(C_\c) = {1\over N_c}  \<0| \Tr  \exp \( i \sigma^a \phi^a \)|0\> \ ,
\label{wI1}\ee
where
\be \phi^a(z_0,\rho) =
{\hbox{\bf R}}^{ab} {\eta^+}^b_{\m\n} \int_{C_\g} \! dx_\m \ (x-z_0)_\n
\vf (x-z_0; \r) \ . \label{iin} \ee We omit the path
ordering operator $\pa$ in Eq. (\ref{wI1}) because the instanton field
(\ref{if1}) is a hedgehog in color space, and so it locks the color
orientation by space coordinates. Although
in certain situations, the integrals of this type (Eq. (\ref{iin})) can be evaluated
explicitly \cc{Sh}, the calculation of the
total integral (\ref{iin}) for a given contour requires an additional
work, so we must restrict
ourselves with the weak-field approximation.
In contrast to our previous paper \cc{DCH}, here we use the cutoff $\l^2$ to
regularize the IR divergences in the instanton case, while the UV
divergences do not appear at all due to the finite instanton size.
Then, in case of the instanton field, the LO contribution reads
\be
W_I^{(1)}(C_\c) = 2 h(\c) \int\! dn(\r) \ \D_1^I(0, \r^2\l^2) \ ,
\ee where
\be
\D_1^I(0, \r^2\l^2) = -  \int \! {d^4 k \over (2\pi)^4} \ex^{ikz}
\d(z^2) \[2 \tilde \vf'(k^2; \r)\]^2\ . \ee
Here, $\tilde \vf (k^2; \r)$
is the Fourier transform of the instanton profile
function $\vf (z^2; \r)$ and $\tilde \vf'(k^2; \r)$ is it's derivative with respect to
$k^2$.
Note, that for the instanton calculations, it is necessary to map
the scattering angle $\c$ to the Euclidean space by the analytical continuation \cc{EUC}
$ \c \to i\g \ $, and perform the inverse transition to the Minkowski space-time in the final
expressions in order to restore the $Q^2$ dependence.
In the singular gauge, when the profile function is
\be
\vf (z^2) = {\r^2 \over z^2 (z^2 + \r^2)} \ ,
\ee one gets
\be
\D_1^I (0, \r^2\l^2) = {\pi^2 \r^4 \over 4} \[\ln (\r^2 \l^2) \ \F_0(\r^2\l^2) + \F_1(\r^2\l^2)
\]\ ,
\ee where
\be
\F_0 (\r^2\l^2) = {1 \over \r^4\l^4} \int_0^1\! {dz \over z (1-z)} \ \[1+ \ex^{\r^2\l^2} -
2 \ex^{z \cdot \r^2\l^2} \]
\ \ , \ \ \lim_{\l^2\to 0}\F_0(\r^2\l^2) = 1 \ 
\ee and
\be
\F_1(\r^2\l^2) = \sum_{n=1}\int_0^1 \! dxdydz \ {[-\r^2\l^2 (xz + y(1-z))]^n \over n!
n}\ex^{\r^2\l^2 [xz + y(1-z)] } \ \ , \ \ \lim_{\l^2\to 0}\F_1(\r^2\l^2) = 0
\
\ee are the IR-finite expressions. At high energy the instanton induced contribution
is reduced to the form
\be
{d \ w_I(Q^2) \over d \ \ln Q^2} = {\pi^2 \over 2}
\dI \ \r^4 \[\ln (\r^2 \l^2) \ \F_0(\r^2\l^2) + \F_1(\r^2\l^2)
\] \equiv - B_I(\l^2)\ .  \label{II1}\ee
Here we used the exponentiation of the single-instanton result
in a dilute instanton ensemble \cc{DCH}:
\be W_I = \exp\(w_I\) \ , \label{eq:exp} \ee
and took only the LO term of the weak-field
expansion (\ref{g1}): $\w1 \to w_I $.

In order to estimate the magnitude of the instanton induced effect
we consider the standard  distribution function \cc{tH} supplied with the
exponential suppressing factor, which has been suggested in Ref. 
\cc{SH2} (and discussed in Ref. \cc{DEMM99} in the framework of constrained instanton
model) in order to describe the lattice data \cc{LAT}
\be
dn(\r) = {d\r \over \r^5} \ C_{N_c} \(2\pi \over \a_s(\m_r) \)^{2N_c} \exp\(- {2\pi \over \a_s (\m_r)}
\) \(\r\m_r\)^{\b} \exp\(- 2 \pi \s \r^2\) \ , \label{dist1}
\ee
where the constant $C_{N_c}$ is
\be
C_{N_c} = {0.466 \ \ex^{-1.679 N_c} \over (N_c-1)! (N_c-2)!}\approx 0.0015 \ ,
\ee
$\s$ is the string tension,  $\b= \b_0+O\[ \a_s(\m_r) \]$, and $\m_r$ is
the normalization point \cc{MOR}. Given the distribution (\ref{dist1})
the main parameters of the instanton liquid model---the mean
instanton size $\bar \r$ and the instanton density $\bar n$---will read
\be
\bar  \r = {\G(\b/2 - 3/2) \over \G(\b/2 - 2)} {1 \over \sqrt{2 \pi \s} } \ , \ee
\be
\bar n = {C_{N_{c}} \G (\b/2 - 2) \over 2} \(2\pi \over \a_S(\bar \r^{-1}) \)^{2N_c}
\({\L_{QCD} \over \sqrt{2\pi \s}}\)^{\b} (2\pi \s)^2 \ . \label{nbar}
\ee In Eq. (\ref{nbar}) we choose, for convenience, the normalization scale
$\m_r$ of order of the instanton inverse mean size $\bar\r^{-1}$.
Note, that these quantities correspond to the mean size $\r_0$ and
density $n_0$ of instantons used in the model \cc{ILM}, where the size distribution
(\ref{dist1}) is approximated by the delta function
$
dn(\r) = n_0 \d(\r-\r_0) d\r \ .$

Thus, we find the leading instanton contribution (\ref{II1}) in the form
\be
B_I=  K \pi^2 \bar n {\bar \rho}^4 \ln{2\pi \s \over \l^2} \[1 + O\({\l^2\over 2\pi\s}\)\]
, \label{pow1} \ee
where
\be
K = {\G(\b_0/2) [\G (\b_0/2-2)]^3 \over 2 \ [\G(\b_0/2-3/2)]^4} \approx
0.74 \ ,
\ee
and we used the one loop expression for the running coupling constant
\be
\a_s (\bar \r^{-1}) = - {2\pi \over \b_0 \ln \ {\bar \r \L}} \ \ ,
 \ \ \b_0 = {11 N_c -
2 n_f \over 3}\ \ . \ee
The packing
fraction $ \pi^2 \bar n {\bar \rho}^4 $ characterizes diluteness
of the instanton liquid and within the conventional picture its value is estimated to be
$ 0.12 \ ,$
if one takes the model parameters as (see Ref. \cc{REV})
\be  {\bar n} \approx 1 \ fm^{-4},
\ \ {\bar \r} \approx 1/3 \ fm \ ,
\ \ \s \approx (0.44 \ GeV)^2. \label{param} \ee
The leading contribution to the quark form factor at asymptotically large $Q^2$
is provided by the (perturbative) evolution governed by
the cusp anomalous dimension (\ref{cp}).
Thus, the instantons yield subleading effects
to the large-$Q^2$ behavior accompanied by a
numerically small factor \be B_I   \approx 0.02 \ , \ee
as compared to the perturbative term
$2C_F /\b_0 \approx 0.24$.

Therefore, from Eqs. (\ref{II1}) and (\ref{npc}),
we find the expression for the quark form factor at large $Q^2$
with the one-loop perturbative contribution and the nonperturbative contributions
(the function $W_{np}$ in Eq. (\ref{npc0})) which include both the IR renormalon
and the instanton induced terms
$$
F_q(Q^2)  $$ \be = \exp\[- {2C_F \over \b_0} \[\ln {Q^2 \over \L^2}
\ln{\ln(Q^2/\L^2) \over \ln(Q_0^2/\L^2)} - \ln {Q^2 \over Q_0^2}
\( 1 -\ln{\ln(Q_0^2/\L^2) \over \ln(\l^2/\L^2)} \) \]  - \ln {Q^2 \over Q_0^2}
\(B_I + \f_{ren}  \) \]
\ .  \label{eq:final} \ee
It is clear, that while the asymptotic (double-logarithmic) behavior is controlled by the
perturbative cusp anomalous dimension, the leading nonperturbative corrections results
in a finite renormalization of the
next-to-leading (logarithmic) perturbative term. From the formal point of view, 
the evolution equation
(\ref{1}) describing the large-$Q^2$ asymptotic is valid even at the low scales
$Q^2 \sim 1 \ GeV^2$, since the only condition 
of applicability of the Wilson integrals approach is
$Q^2 >> \l^2, m^2$. However, in the low-energy domain the 
perturbative one-loop cusp anomalous dimension
$\G_{cusp} (\a_s)$ (\ref{cp}) should be supplemented by higher loop corrections, and thus the
explicit formula (\ref{eq:final}) would include additional logarithmic terms. The relevance of the
instanton induced part (\ref{II1}) in the low-energy domain calculated in the dilute gas 
approximation
can be questioned, and the additional
consideration within more proper framework may help one to verify it. Indeed, the corrections to the
single instanton approximation may be large at sufficiently low momenta (see for recent discussions,
{\it e.g.} Ref. \cite{Faccioli:2002wr} and references therein).
At the moment, we can say confidently
that the evolution equation is valid at $Q^2 \geq 1 \ GeV^2$.

We have to comment that the weak field limit used in the instanton calculations
may deviate from the exact result. Nevertheless, we expect that using of the instanton
solution in the singular gauge, that concentrate the field at small distances,
leads to the reasonable numerical estimate of the full effect.
Thus, the resulting diminishing of the instanton contributions with respect
to the perturbative result appears to be reasonable output. It should be emphasized that
in the present paper, all the calculations have been performed analytically while the evaluation
of the instanton contribution beyond the weak field approximation requires a numerical
analysis, which will be the subject of a separate work. Moreover, the use of the singular gauge for
the instanton solution allows us to prove the exponentiation theorem for the Wilson loop in
the instanton field \cc{DCH} which permits one to express the 
full instanton contribution as the exponent
of the all-order single instanton result (\ref{eq:exp}).

\section{Conclusion}

We analyzed the structure of the nonperturbative corrections to the quark
form factor at large momentum transfer. In order to model the nonperturbative effects,
we studied the quark scattering process in the background of the instanton vacuum.
The instanton induced contribution to the color singlet quark form factor is calculated in
the large momentum transfer regime. It was shown that the instanton induced corrections
correspond to the leading term proportional to $\ln Q^2$. The magnitude of these corrections
is determined by the small instanton liquid packing fraction parameter,
and they can be treated as finite renormalization of the subleading perturbative
part (\ref{eq:final}). In addition to this, the minimal set of the nonperturbative parameters
is found considering infrared renormalon ambiguities of the perturbative series. Within this approach,
it is shown that the leading large-$Q^2$ behavior of nonperturbative contributions should also
be determined by the logarithmic term $\sim \ln \ Q^2$, what is consistent with the instanton analysis.

Let us emphasize that our results are quite sensitive to the prescription how to  make the integration
over instanton sizes finite. For example, if one used the sharp cutoff then the instanton
would produce strongly suppressed power corrections such as $\propto (\L/Q)^{\beta_0}$.
However, we think that the distribution function (\ref{dist1}) should be considered
as more realistic, since it reflects more properly the structure of
the instanton ensemble modeling the QCD vacuum.
Indeed, this shape of distribution was recently advocated in Refs. \cc{SH2, DEMM99} and supported by the lattice
calculations \cc{LAT}.

Finally, we think that the instanton induced effects are more interesting for theoretical investigation
and more important
for phenomenology in the hadronic processes which possess two energy scales, such as the
total center-of-mass energy
$s$ (hard characteristic scale), and the squared momentum transfer $-t$ which is small compared
to the latter: $-t << s$. One of the most interesting examples of such processes is the parton-parton scattering
and the soft Pomeron problem \cc{Sh, KKL, KRP}. Another important situations where the nonperturbative
(including instanton induced) effects can emerge are the transverse momentum distribution
of vector bosons in the Drell-Yan process (see, {\it e.g.,} Refs. 
\cc{KRREN, TAF}), and the phenomenon of saturation
in deep-inelastic scattering (DIS) at small x \cc{SCHUT}. 
The explicit evaluation of the instanton effects
in these processes will be the subject of our forthcoming study.

\section{ Acknowledgments}
\noindent
We are grateful to A. P. Bakulev, N. I. Kochelev,  O. V. Teryaev
and especially to S. V. Mikhailov for useful discussions.
This work was partially supported by RFBR (Grant Nos. 03-02-17291, 02-02-81023, 02-02-16194,
01-02-16431), the Heisenberg-Landau program (Grant No. HL-2002-13), the Russian Federation 
President's Grant No. 1450-2003-2, 
and INTAS (Grant no. 00-00-366). One of the authors (I.Ch.) is grateful to
the Theory group at DESY (Hamburg), and especially to F. Schrempp,  
for invitation, financial support, and
numerous fruitful discussions.

\end{document}